\documentclass[journal]{ijdatics}
\usepackage{cite}
\usepackage{color}
\usepackage{graphicx}

\ifCLASSINFOpdf
\else
\fi
\begin{document}
\setcounter{page}{22}

\title{A Novel VSWR-Protected and Controllable CMOS Class E Power Amplifier for Bluetooth Applications}

\author{Wei~Chen,~\IEEEmembership{}
        Wei~Lin,~\IEEEmembership{}
        and~Shizhen~Huang~\IEEEmembership{}
\thanks{All authors are with the Fujian Key Laboratory of
 Microelectronics $\&$ Integrated Circuits, Fuzhou University, Fujian Province, 350002, China PRC. E-mail: wchen@fzu.edu.cn.}
}

\markboth{International Journal of Design, Analysis and Tools for Circuits and Systems,~Vol.~1, No.~1, June~2011}%
{Chen \MakeLowercase{\textit{et al.}}: An IJDATICS Article}
\maketitle

\begin{abstract}
This paper describes the design of a differential class-E PA for
Bluetooth applications in 0.18$\mu$m CMOS technology with load mismatch
protection and power control features. The breakdown induced by load
mismatch can be avoided by attenuating the RF power to the final
stage during over voltage conditions. Power control is realized by
means of ``open loop'' techniques to regulate the power supply
voltage, and a novel controllable bias network with temperature
compensated is proposed, which allows a moderate power control slope
(dB/V) to be achieved. Post-layout Simulation results show that the
level of output power can be controlled in 2dBm steps; especially
the output power in every step is quite insensitive to temperature
variations.
\end{abstract}

\IEEEpeerreviewmaketitle

\section{Introduction}
\IEEEPARstart{B}{luetooth} devices operate in the 2400-2483.5MHz
Industrial, Scientific and Medical (ISM) band. There are basically
three classes based on the transmission distance. They are Class 1
(The transmitted output power is 20dBm), Class 2 (The transmitted
output power is 4dBm) and Class 3 (The transmitted output power is
0dBm) respectively. Usually, the Bluetooth power amplifier is
working in low power model, so the output power of Class 1 power
amplifier must controllable down to 4dBm or less in a monotonic
sequence to save the power~\cite{VSL:01}.

A standard method of controlling the output power of a power
amplifier is to use a voltage regulator to regulate the battery or
power supply voltage. Typical approaches to controlling the output
power of a power amplifier use an ``open loop'' or a ``closed loop''
control technique. ``Closed loop'' techniques use an RF sensor, such
as a directional coupler, to detect the power amplifier output
power. The detected output power is used in a feedback loop to
regulate the output power. ``Open loop'' techniques control the
output power by regulating either the power supply voltage or power
supply current used by the power amplifier. ``Open loop'' techniques
are popular since open loop techniques do not have the loss and
complexity associated with RF sensor elements. But in conventional
power control schemes by mean of regulating only power supply
voltage, the PA gain control slope (dB/V) is precipitate and the PA
will suffer from transmit burst shaping and potential stability
problems.

Nowadays, Gallium Arsenide (GaAs), BiCMOS and silicon bipolar
technologies still dominate in the power amplifier design. Compared
with CMOS technology, these technologies offer higher breakdown
voltage, lower substrate loss and higher quality of monolithic
inductors and capacitors, but they are expensive. CMOS technology,
on the other hand, could provide single-chip solution which greatly
reduces the cost. But CMOS technology suffers from poor quality
factors of monolithic passive components, low breakdown voltage of
the transistors and large process variation.

More still, the main obstacle to the actual exploitation of  CMOS
PAs is the ruggedness requirement, i.e., the ability to survive
under high load voltage standing wave ratio (VSWR) conditions with a
full-power RF drive~\cite{SCP:04}. Typically, device testing procedures
for commercial PAs can demand a VSWR as high as 10:1 under a 5V
power supply. Such a strong mismatch condition results in very high
voltage peaks at the collector of the final stage (much higher than
the nominal supply voltage) and may eventually lead to permanent
failure of the power transistor due to avalanche breakdown. As
reported in~\cite{Y:00}, to comply with ruggedness requirements,
collector voltage peaks in excess of 16V have to be tolerated. CMOS
transistors usually exhibit lower breakdown voltages. Thick
gate-oxide transistors of TSMC 0.18$\mu$m RF CMOS process have a
6.8V breakdown voltage.

In this paper, a two-stage 0.18$\mu$m CMOS monolithic PA for Class 1
Bluetooth is proposed. The PA includes a power control circuit which
can improve the PA gain control slope (dB/V) and a protection
circuit to overcome the detrimental effects of load impedance
mismatch. The level of output power can be controlled in 2dB steps
using an open loop control technique and a novel linearity control
bias network using temperature compensated, and achieved confirm
that ruggedness specifications can be fulfilled.

\section{Circuits design}
The power amplifier is designed in a 0.18$\mu$m CMOS technology with
analog and RF options. This CMOS technology has two kinds of
transistors. Thick gate-oxide transistors, which are similar to
0.35$\mu$m transistors, have a higher breakdown voltage. Thin
gate-oxide transistors that are similar to 0.18$\mu$m transistors
have a higher Gm. So the thin gate-oxide transistors are chosen in
driver stage to generate a larger signal to turn the transistor on
and off. And the thick gate-oxide transistors used for out stage.
\begin{figure}[htb]
\centerline{
\includegraphics[width=7.5cm]{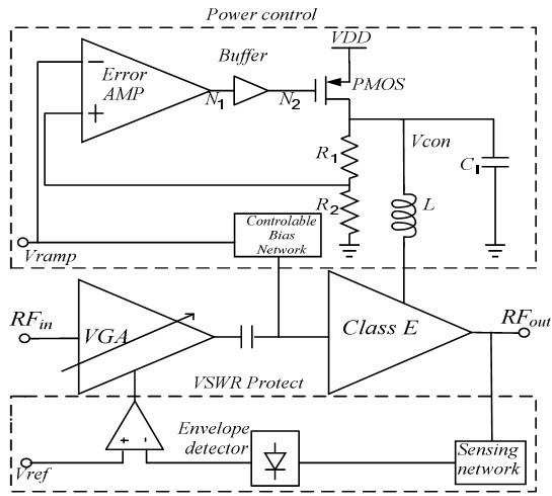}}
\caption{Simplified power amplifier topology with power control and
VSWR protection.}
\end{figure}
Fig.1 shows a simplified power amplifier topology which contains
three main modules, Class E power amplifier and Driver stag; Power
control module and VSWR protect module.
\subsection{Class E Power Amplifier  and Driver Stage For Bluetooth}
Class E power amplifier is a switching-mode amplifier, which is nonlinear
amplifier that achieves efficiencies approaching 100
To achieve these conditions, all the components should be properly
designed. As shown in Fig.2, the loading
\begin{figure}[htb]
\centerline{
\includegraphics[width=7.5cm]{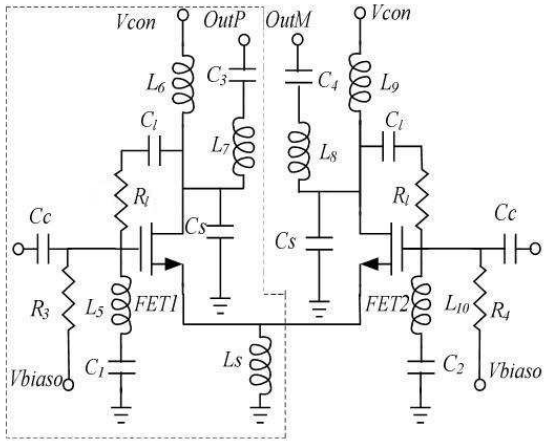}}
\caption{Differential class E PA with finite ground inductance.}
\end{figure}
inductor $L_6$ is either a RF choke (RFC) or a finite inductance. Cs
is a charging capacitor; $L_7$ and $C_3$ are designed to be a series
\emph{LC} resonator with an excess inductance at the frequency of
interest. The resonator resonates at the fundamental frequency, and
suppresses the other harmonics. The purpose of \emph{LC} resonator
is designed for optimization conditions. The optimum values for each
component are calculated as follows~\cite{SS:75}.
\begin{equation}
R_l  = \frac{{0.577(V_{dc} - V_{knee} )^2 }}{{P_{out} }}
\end{equation}
\begin{equation}
C_3  = \frac{1}{{5.447\omega R_l }}
\end{equation}
\begin{equation}
L_7  = \frac{{QR_l }}{\omega }
\end{equation}
\begin{equation}
C_s  = C_3 \times\frac{{5.447}}{Q}(1 + \frac{{1.42}}{{Q - 2.08}})
\end{equation}
where $R_l$ is the optimize load, Q is the quality factor of
\emph{LC} resonator.
\begin{figure}[htb]
\centerline{
\includegraphics[width=7.5cm]{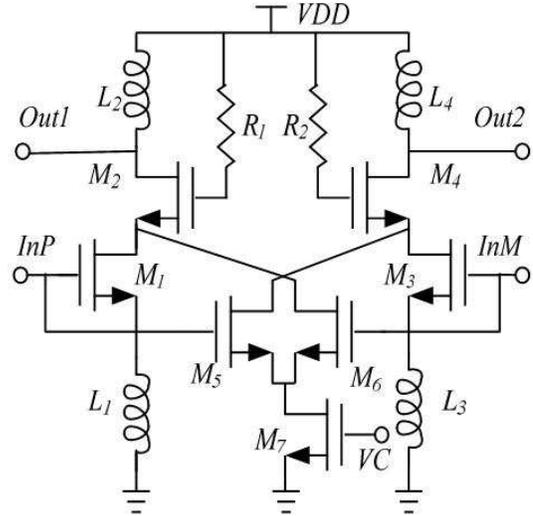}}
\caption{Drive stage in a cascade topology.}
\end{figure}
Fig.3 shows a simplified circuit of drive stage in a cascade
topology, which is to generate a larger signal to turn the
transistor on and off. The variable-gain amplifier (VGA), which is
operated at maximum gain under nominal conditions When \emph{VC} is
in low voltage and $M_7$ turns off. When $M_7$ turns on, the gain of
VGA will decrease, and attenuate the RF power to the final stage and
the drain voltage of final transistors will decrease.

\subsection{Open-loop Power Control for Class E Power Amplifier}
The open-loop control system in which the output has no effect upon
the input signal.The methodology to realize power controllable is to
change the output stage power supply voltage in ``open loop" control
technique which is a likely LDO and to change the final stage bias
by a novel controllable network temperature compensated. The benefit
of using this topology is that the noise of its output voltage is
lower and the response to input voltage transient and output load
transient is faster. The output voltage \emph{$V_{con}$} in Fig.1
can obtain from equation (5).
\begin{equation}
Vcon = (1 + \frac{{R1}}{{R2}}) \times V_{ramp}
\end{equation}
When the Bias of Class E is in a fixed station, the drive signal is
an excellent switch signal, and not to consider other non ideal
factors the output power \emph{$Pout$} can be obtained from equation
(6).
\begin{equation}
Pout = 0.577\times\frac{{Vcon^2 }}{{Rload}} = 0.577\times[(1 +
\frac{{R1}}{{R2}}) \times V_{ramp}]^2 /Rload
\end{equation}
Equation (6) shows that the output power \emph{$Pout$} and the
control signal \emph{$V_{ramp}$} are in a square relationship.
Actually to change the driven signal can improve PA gain control
slope (dB/V).
\begin{figure}[htb]
\centerline{
\includegraphics[width=7.5cm]{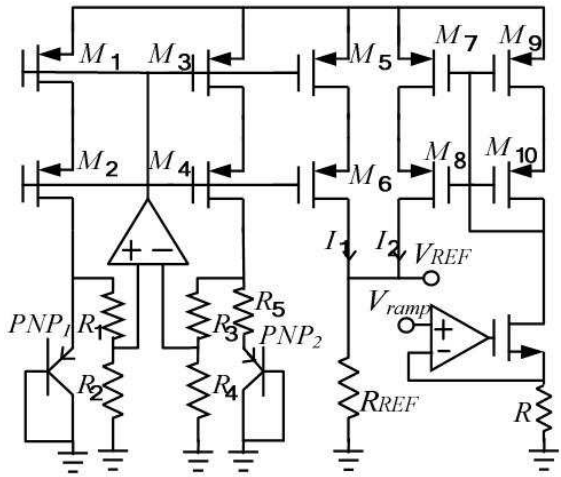}}
\caption{Novel linearity controllable bias network with temperature
compensated.}
\end{figure}
Fig.4 shows a simplified novel linearity controllable bias network
with temperature compensated for final stage. The output voltage can
be obtained from equation (7)
\begin{equation}
V_{REF}  = a + a_1 \times V_{ramp}
\end{equation}
It shows \emph{$V_{REF}$} is controlled by \emph{$V_{ramp}$}, Where
\[
a = I_1 \times R_{REF}  = (\frac{{V_{BE1} \times R_{REF} }}{{R_3  + R_4 }}) +
(\frac{{V_T \ln n \times R_{REF} }}{{R_5 }})
\]
\[
a_1  = \frac{{I_2 \times R_{REF} }}{{V_{ramp} }} = \frac{{R_{REF}
}}{R}\times (\frac{{W_9 }}{{L_9 }}/\frac{{W_7 }}{{L_7 }})
\]
In normal temperature,
\[
\frac{{\partial V_{BE} }}{{\partial T}} \approx  - 1.5mV/^\circ
K,\frac{{\partial V_T }}{{\partial T}} \approx  + 0.087mV/^\circ K,
\]
\[
\frac{{\partial a}}{{\partial T}} = (\frac{{R_{REF} }}{{R_3  + R_4
}}\times \frac{{\partial V_{BE1} }}{{\partial T}}) + (\frac{{\ln n \times R_{REF}
}}{{R_5 }} \times \frac{{\partial V_T }}{{\partial T}})
\]
The coefficient ``$a$'' and ``$a_1$'' are constant, but ``$a$'' is
proportional to temperature, and ``$a_1$'' insensitive to
temperature.

\subsection{Closed-Loop Drain Peak Voltage Control}
There is no isolator used between the PA and the antenna as a result
the power amplifier can cause strong load mismatch due to faults or
disconnection antenna. Therefore power transistors should be able to
tolerate over voltage, as the peaks of drain voltage waveforms show
much higher under mismatch conditions than under nominal conditions.
The worst case conditions occur when the power amplifier is operated
under both oversupply and load mismatch conditions. So the
ruggedness specification is usually considered in terms of a maximum
tolerable output VSWR under a specified oversupply condition.
Typical data sheets of commercial power amplifiers guarantee that no
permanent damage happened with 10:1 load VSWR under supply voltage
of 5V.

The risk of breakdown can be prevented by simply attenuating the RF
signal which drives the final stage during over voltage conditions~\cite{SCCP:03,BSMR:06}. This can be achieved by adopting a feedback control
system, which detects the peak voltage at the output collector node
and decreases its value to a specific threshold by varying the
circuit gain.

\begin{figure}[htb]
\centerline{
\includegraphics[width=7.5cm]{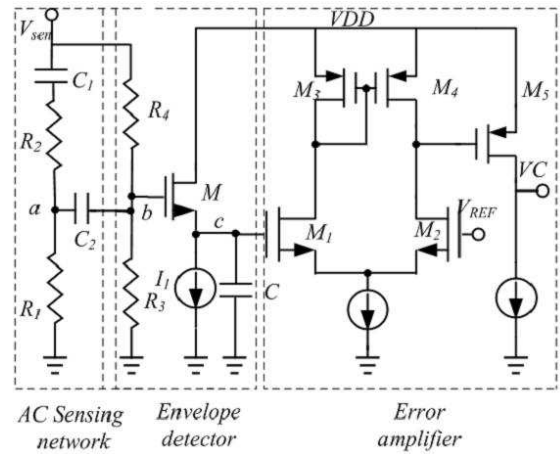}}
\caption{Closed-loop drain peak voltage control.}
\end{figure}

Drain voltage of the output transistor is scaled down by a
high-input-impedance sensing network, the scaled down voltage is
applied to an envelope detector delivering an output signal
proportional to the collector peak voltage. An error amplifier then
compares the rectified waveform voltage with a reference voltage.
Finally, the output error is used to control the gain of the drive
stage.

If $ R_3 (R_4 ) >  > R_2$
 and capacitors $C_1$ and $C_2$ are large enough to be considered short
 circuits at the operating frequency, then the output voltage of
 rectifier (at node c) can be expressed as
\small
\begin{equation}
V_c  = (V_{sen(peak)}  - Vcon) \times \frac{{R_1 }}{{R_1  + R_2 }} +
Vcon \times \frac{{R_3 }}{{R_3  + R_4 }} - Vgs(M)
\end{equation}
\normalsize
where \emph{Vgs(M)} is the voltage between the gain and the source
of transistor M,\emph{$V_{sen(peak)}$} is the peak of the drain
voltage of final stage transistor FET1, and \emph{$V_{con}$} is the
DC voltage of it. When the drain voltage of final stage transistor
over the reconverted voltage, then \emph{$V_c$} will exceed
\emph{$V_{REF}$}, and the VSWR protection will in work station.

\section{Layout design}
The power amplifier was first layered out using Cadence Analog
Artist, and then imported into ADS's Momentum RF 2.5d
electromagnetic simulator. Pins are added to layout to define the
current flow direction, the polygons are meshed into rectangles and
triangles, and the dielectric properties of the substrate are
defined. Fig.6 shows layout of differential class E with on chip
input and inter-stage matching. The circuit is then simulated using
the planar field solver.
\begin{figure}[htb]
\centerline{
\includegraphics[width=7.5cm]{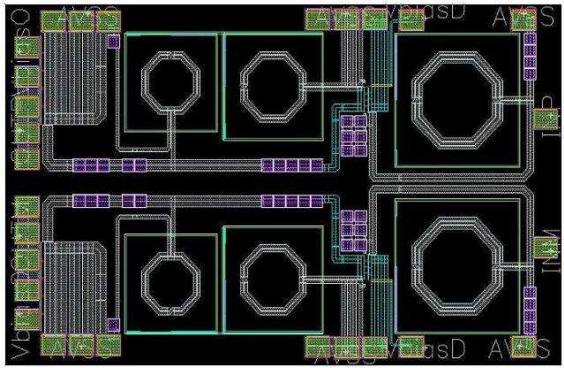}}
\caption{Layout of differential class E with on chip input and
inter-stage matching.}
\end{figure}
The layouts of RF devices, especially for power amplifiers, require
special attention. The output transistor carries 250mA of dc
current, plus the RF current, and out stage transistors ($M_8$ and
$M_9$) has a total width of 2.4mm. The drain contact area of each
transistor is enlarged, and parallel layers of metal 1 to metal 5
are used as drain and source connections such that the device is
able to handle large currents. The output devices are placed as
close as possible to the output pads. Many bond-wires can handle the
large output currents.14 ground pads were used in order to minimize
the ground bond-wire inductance.

\section{Simulation results}
The PA in Fig. 1 was simulated and optimized by using ADS (Advanced
Design System) in 0.18$\mu$m technology, and the Bond wire
inductance is replaced by a physical lumped element model.
\begin{figure}[htb]
\centerline{
\includegraphics[width=7.5cm]{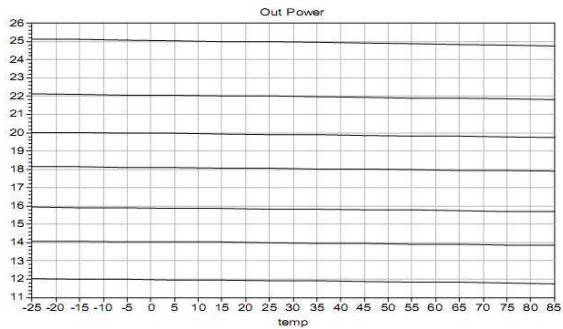}}
\caption{Output power vary with temperature.}
\end{figure}
Fig.7 shows the output power (Spectrum) vary with Temperature,
it indicates that the output power changes less than 0.3dBm when the
temperature vary from -25$^{\circ}\mathrm{C}$ to
85$^{\circ}\mathrm{C}$.
\begin{figure}[htb]
\centerline{
\includegraphics[width=7.5cm]{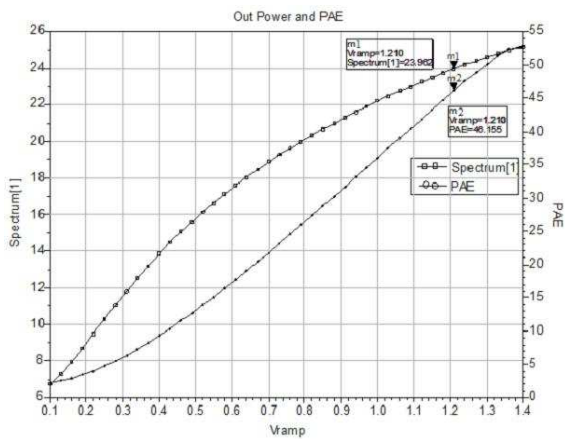}}
\caption{The output power and PAE vary with control voltage.}
\end{figure}
Fig.8 shows the PAE and Output power (Spectrum) vary with the
control voltage \emph{$V_{ramp}$}. When used 0dBm input signal and
1.8V supply voltage at 2.45GHz the PA reached to the maximum output
power of 25.1dBm and 54.2$\%$ power-added efficiency (PAE).
\begin{figure}[htb]
\centerline{
\includegraphics[width=7.5cm]{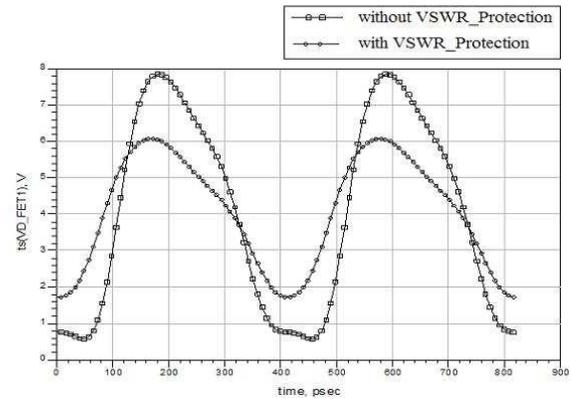}}
\caption{Drain voltage of FET1 under over power supply and load
mismatch.}
\end{figure}
Fig.9 shows the drain voltage of FET1 would reach much higher more
than 6.8V without VSWR protection when the supply voltage is 5V and
the load resistance is 5$\Omega$, then the transistor FET1 will
breakdown. But the transistor will be in safe station with VSWR
protection.

\section{Conclusion}
A two stage of class-E power amplifier for class 1 Bluetooth
applications has been designed which includes a protection circuit
preventing output stage failure due to severe load mismatch. Safe
operation was achieved through the use of a feedback loop that acts
on the circuit gain to limit the overdrive of the output transistor
whenever an over voltage condition is detected. The output power can
be controlled easily by a variable supply implemented by ``open
loop'' technique; also a novel bias network controlled by
``$V_{ramp}$'' with temperature compensated for final stage is
proposed which allows a moderate power control slope (dB/V) to be
achieved. Post-layout simulation a 25.1dBm output power and 54.2$\%$
PAE were achieved at a nominal 1.8V supply voltage. The amplifier is
able to sustain a load VSWR as high as 10:1 up to a 5V supply
voltage without exceeding the breakdown limits. And the level of
output power can be controlled in 2dBm steps; especially the output
power in every step is quite insensitive to temperature variations.
And it is satisfied for Bluetooth applications.

\section*{Acknowledgment}
The authors would like to thank the teachers in Fujian key
Laboratory of Microelectronics $\&$ Integrated Circuits, they are
very kind and patient, and would like to thank Fujian Integrated
Circuit Design Center for the use of their facilities. The project
was supported by the Natural Science Foundation of China (Grant No.
10871221).


%

\ifCLASSOPTIONcaptionsoff
  \newpage
\fi

%

\begin{IEEEbiography}
[{\includegraphics[width=1in,height=1.25in,clip,keepaspectratio]{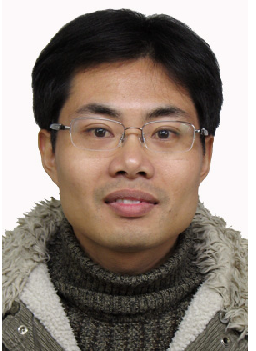}}]{Wei~Chen}
was born in Putian, Fujian in 1977. He received M.S. degree in
College of Physics and Information Engineering from Fuzhou
University in 2007. He is currently employed by Physics and Information
Engineering, Fuzhou University. His current interests include analog
integrated circuits design and sensor technology research.
~\\~\\~\\~\\~\\~\\~\\~\\~\\~\\~\\~\\~\\~\\~\\~\\~\\~\\~\\~\\~\\~\\~\\~\\~\\~\\~\\~\\~\\~\\~\\~\\~\\~\\~\\~\\~\\~\\~\\~\\~\\~\\
\end{IEEEbiography}

\newpage

\begin{IEEEbiography}
[{\includegraphics[width=1in,height=1.25in,clip,keepaspectratio]{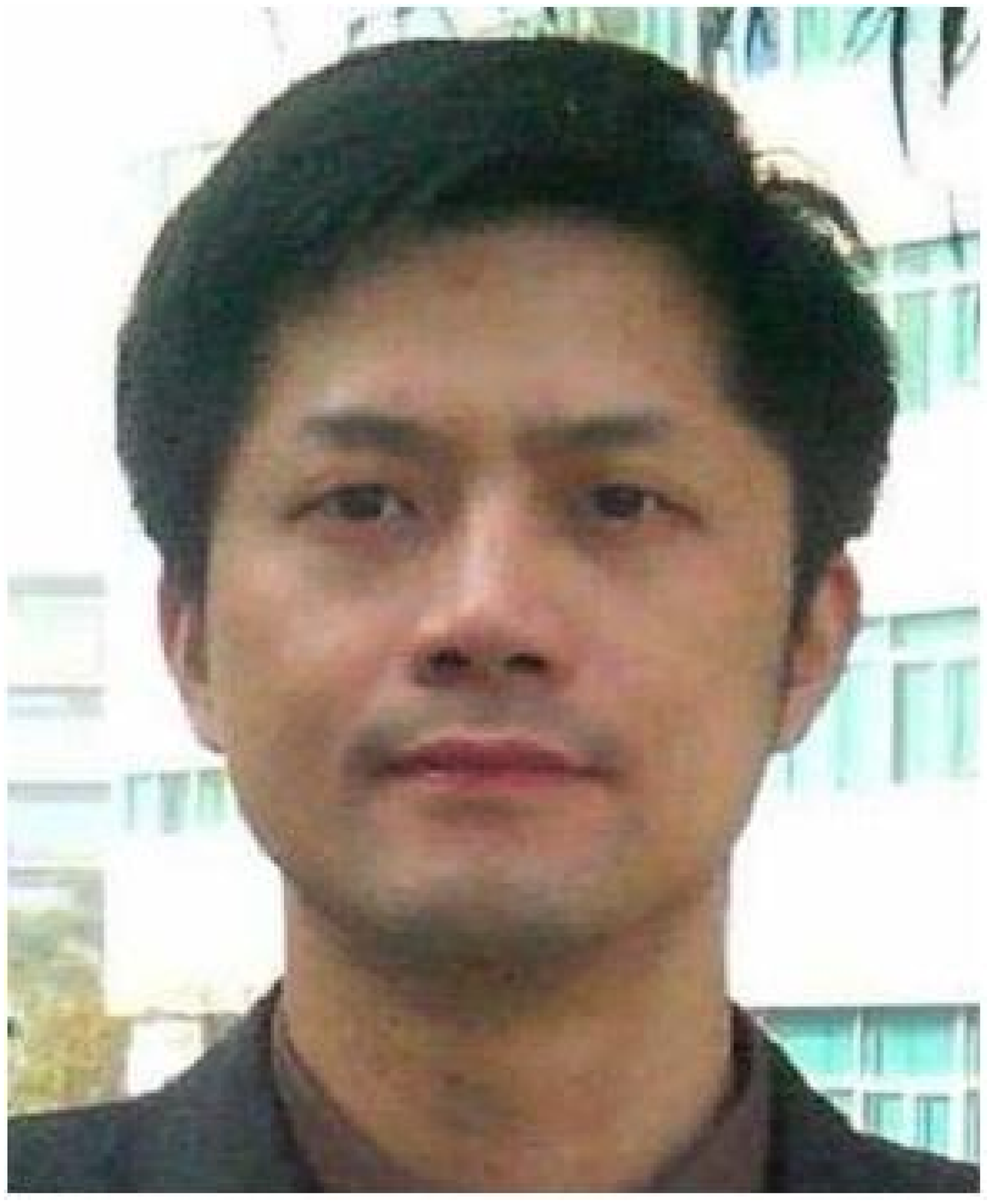}}]{Wei~Lin}
was born in Fuzhou, Fujian in 1968. He received M.S. degree from
Fuzhou University in 1998. He is currently employed by Physics and
Information Engineering, Fuzhou University. His current interests
include analog integrated circuits design and sensor technology
research.

\end{IEEEbiography}

\begin{IEEEbiography}[{\includegraphics[width=1in,height=1.25in,clip,keepaspectratio]{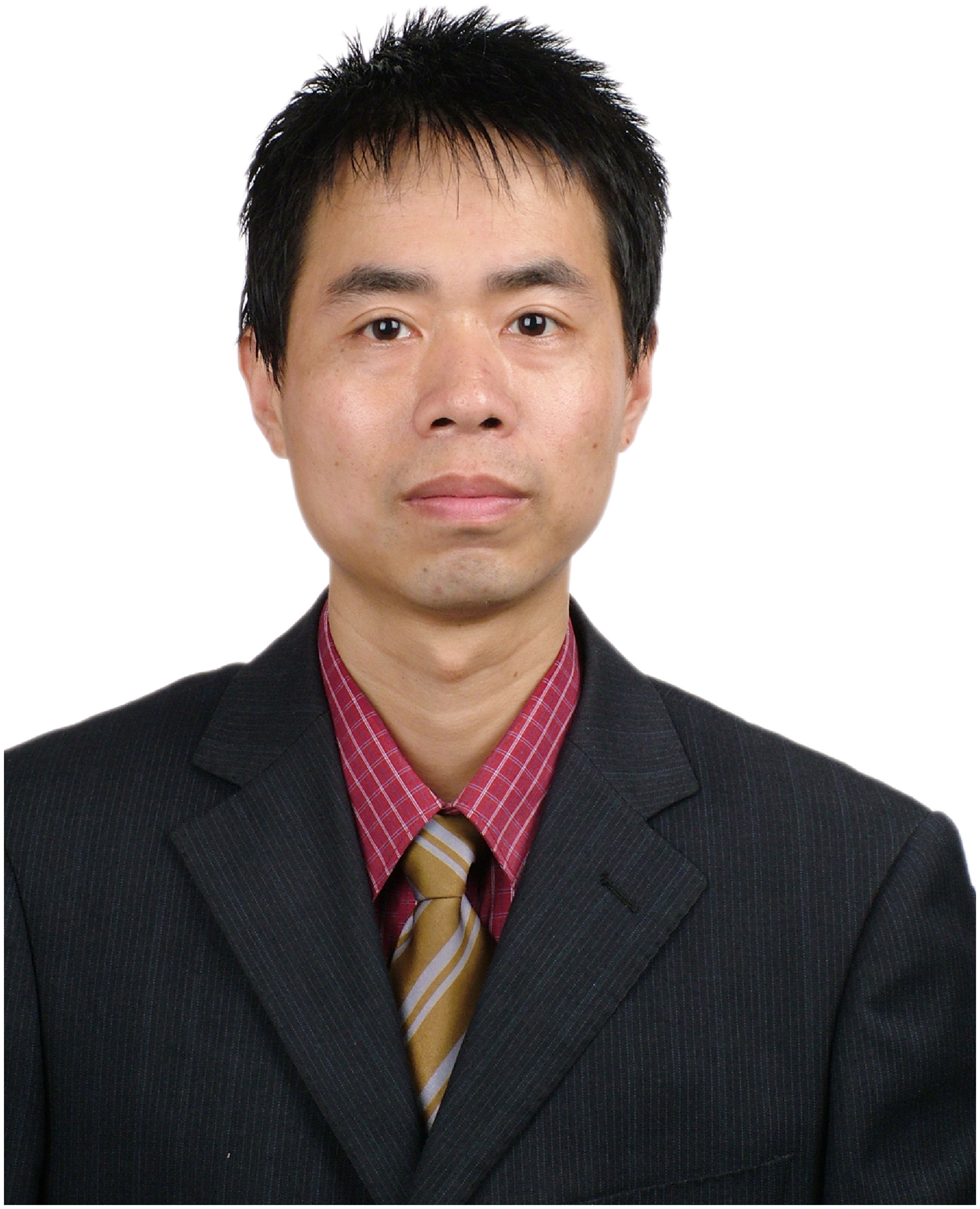}}]{Shizhen~Huang}
was born in Fujian in 1968. He received M.S. degree from Fuzhou
University in 2002. He is currently employed by Physics and Information
Engineering, Fuzhou University. His current interests include analog
integrated circuits design and sensor technology research.
~\\~\\~\\~\\~\\~\\~\\~\\~\\~\\~\\~\\~\\~\\~\\~\\~\\~\\~\\~\\~\\~\\~\\~\\~\\~\\~\\~\\~\\~\\~\\
\end{IEEEbiography}




\end{document}